\documentclass[9pt,twocolumn,twoside]{osajnl}

\journal{josab} 

\setboolean{shortarticle}{true} 

\ifthenelse{\boolean{shortarticle}}{\colorlet{color2}{color2b}}{\colorlet{color2}{color2}} 

\title{A brief overview of the interplay between nonlinearity and transverse Anderson localization}
\usepackage{graphicx}
\author[1,2,*]{Arash Mafi}

\affil[1]{Department of Physics \& Astronomy, University of New Mexico, Albuquerque, NM 87131, USA}
\affil[2]{Center for High Technology Materials, University of New Mexico, Albuquerque, NM 87106, USA}

\affil[*]{Corresponding author: mafi@unm.edu}
\dates{Compiled \today}

\ociscodes{Anderson localization; Transverse Anderson localization; Disorder; Nonlinearity; Optical waveguide; Optical fiber;
Disordered optical fiber; Anderson localizing optical fiber.}

\doi{}

\begin{abstract}
This article presents a brief overview of the interplay between nonlinearity and Anderson localization 
of light, primarily in the context of transverse Anderson localization. The focus is on 
whether Anderson localization is preserved, enhanced, weakened, or destroyed in the presence of nonlinearity. 
Several recent experimental and theoretical results are highlighted and reviewed. 
\end{abstract}

\setboolean{displaycopyright}{true}

\def\NLSE{nonlinear Schr\"{o}dinger equation}

\begin{document}

\maketitle
\thispagestyle{fancy}

\ifthenelse{\boolean{shortarticle}}{\ifthenelse{\boolean{singlecolumn}}{\abscontentformatted}{\abscontent}}{}

\section{Introduction}

Is Anderson localization preserved, enhanced, weakened, or destroyed in the presence of nonlinearity? 
In this short article, we will learn that there may not be a straightforward answer to this question.
A survey of the literature published on this subject reveals that the answer to this rather basic question
can best be described as inconclusive at this point. Of course, it should not be a surprise that the
potentially chaotic behavior as a result of the nonlinear dynamics is largely behind the present state of debate
in the literature on the interplay between disorder and nonlinearity in the context of Anderson localization.
 
This article presents a survey of some highlights in the literature over the past few years with the hope that 
an eventual conclusive answer can be found to such a fundamental question in a not so distant future. The 
question can be cast in a limited form dealing only with the optical Kerr effect, which is of greater
interest to the photonics community, or in a more general context to explore and differentiate the impact of various forms of 
nonlinearity on localization. The interplay between disorder and nonlinearity has been explored over 
the years in various contexts and different disciplines. Some inevitable overlap remains in the existing literature
belonging to different fields of study, primarily due to similarities among the dynamical equations of different 
physical systems. For example, the Gross--Pitaevskii equation describing the Bose-Einstein condensate is 
similar to the \NLSE~describing the propagation of light in a nonlinear optical medium~\cite{BoydBook,AgrawalBook}.

Anderson localization in general, and even Anderson localization of light in particular are broad research topics. In the optical
domain, the most conclusive results on localization have so far been obtained in the context of transverse Anderson localization of light,
where light is transversely localized in a waveguide-like geometry with a disordered transverse profile and 
propagates freely in a longitudinally uniform medium. Anderson localization is more readily observed in this transverse form 
as explained in the next section. Moreover, the transversely localized and longitudinally propagating platform allows for a
relatively long-distance interaction of light with the optical medium, resulting in an appreciable nonlinear phase shift. 
As such, we will primarily focus on results that are more relevant to the impact of nonlinearity on transverse Anderson localization. 
For more details, we refer the interested reader to an excellent review article on this subject by Fishman, Krivolapov, and 
Soffer, in Ref.~\cite{Fishman2012} and references therein.

\section{Anderson localization of light}

Anderson localization is the absence of diffusive wave transport in highly disordered 
scattering media~\cite{Anderson1958,John1987,Sheng2006,Segev2013}. Its origin dates back to a theoretical 
study conducted by Anderson ~\cite{Anderson1958} who investigated the behavior of spin diffusion and 
electronic conduction in random lattices. It was soon realized that because the novel 
localization phenomenon is due to the wave nature of the quantum mechanical electrons
scattering in a disordered lattice, it can also be observed in other coherent wave systems, 
including classical ones~\cite{John1991,Anderson1985,Abrahams2010,Lagendijk2009}. 

The fact that Anderson localization was deemed possible in non-electronic systems was
encouraging, given that the observation of disorder-induced localization was shown to be inhibited
by thermal fluctuations and nonlinear effects in electronic systems. Subsequently,
localization was studied in various classical systems including in acoustics, elastics,
electromagnetics, and optics~\cite{John1991,Lagendijk2009,Graham1990,Hu2008,Chabanov2000}. It was also recently investigated in 
various quantum optical systems, such as atomic lattices~\cite{Billy2008} and propagating photons~\cite{Abouraddy2012}. 

General studies of Anderson localization have revealed that waves in one-dimensional (1D) and 
two-dimensional (2D) unbounded disordered systems are always localized. However, in order for 
a three-dimensional (3D) random wave system to localize, the scattering strength needs to be larger 
than a threshold value. This statement is often cast in the form of $kl^\ast\sim 1$, where $k$ is the effective 
wavevector in the medium, and $l^\ast$ is the wave scattering transport length. This is referred 
to as the Ioffe-Regel condition~\cite{IoffeRegel} and shows that in order to observe Anderson 
localization, the disorder must be strong enough such that the wave scattering transport length 
becomes on the order of the wavelength. The Ioffe-Regel condition is notoriously difficult to
satisfy in 3D disordered optical media. For the optical field
to localize in 3D, very large refractive index contrasts are required that are not generally
available in low-loss optical materials~\cite{John1991}. {\em The fact that Anderson localization is hard to achieve 
in 3D optical systems may be a blessing in disguise; otherwise, no sunlight would reach the earth 
on highly cloudy days.} 

\subsection{Transverse Anderson localization of light}

Unlike 3D  lightwave systems in which the observation of the localization is prohibitively difficult, 
observation of Anderson localization in a quasi-2D optical system is readily possible, as was first 
shown by Abdullaev \textit {et al}.~\cite{Abdullaev1980} and De~Raedt \textit {et al}.~\cite{DeRaedt1989}.
In particular, De~Raedt \textit {et al}.~\cite{DeRaedt1989} showed that 2D Anderson localization can be observed in 
a dielectric with transversely random and longitudinally uniform refractive index profile~\cite{Mafi-AL-AOP-2015}. 
An optical field that is launched in the longitudinal direction tends to remain localized in the transverse plane 
as it propagates in the longitudinal direction in a transversely random dielectric medium. 
This behavior was dubbed transverse Anderson localization of light.

In their pioneering work, Schwartz \textit {et al}.~\cite{Schwartz2007} wrote the transversely 
disordered and longitudinally invariant refractive index profiles in a photorefractive
crystal using a laser beam. They used another probe beam to investigate 
the transverse localization behavior. Their experiment was quite interesting as it allowed them to vary 
the disorder level by controlling the laser illumination of the photorefractive crystal in a controlled
fashion to observe the onset of the transverse localization and the change in the localization 
radius as a function of the disorder level. The transverse localization of the beam and the free 
longitudinal propagation due to the longitudinal invariance of the dielectric medium strongly resembled
the optical waveguides; therefore, applications of Anderson localization for light propagation in an optical fiber-like
medium seemed an appealing extension of these ideas. 

\begin{figure}[h]
\centering\includegraphics[width=3.5in]{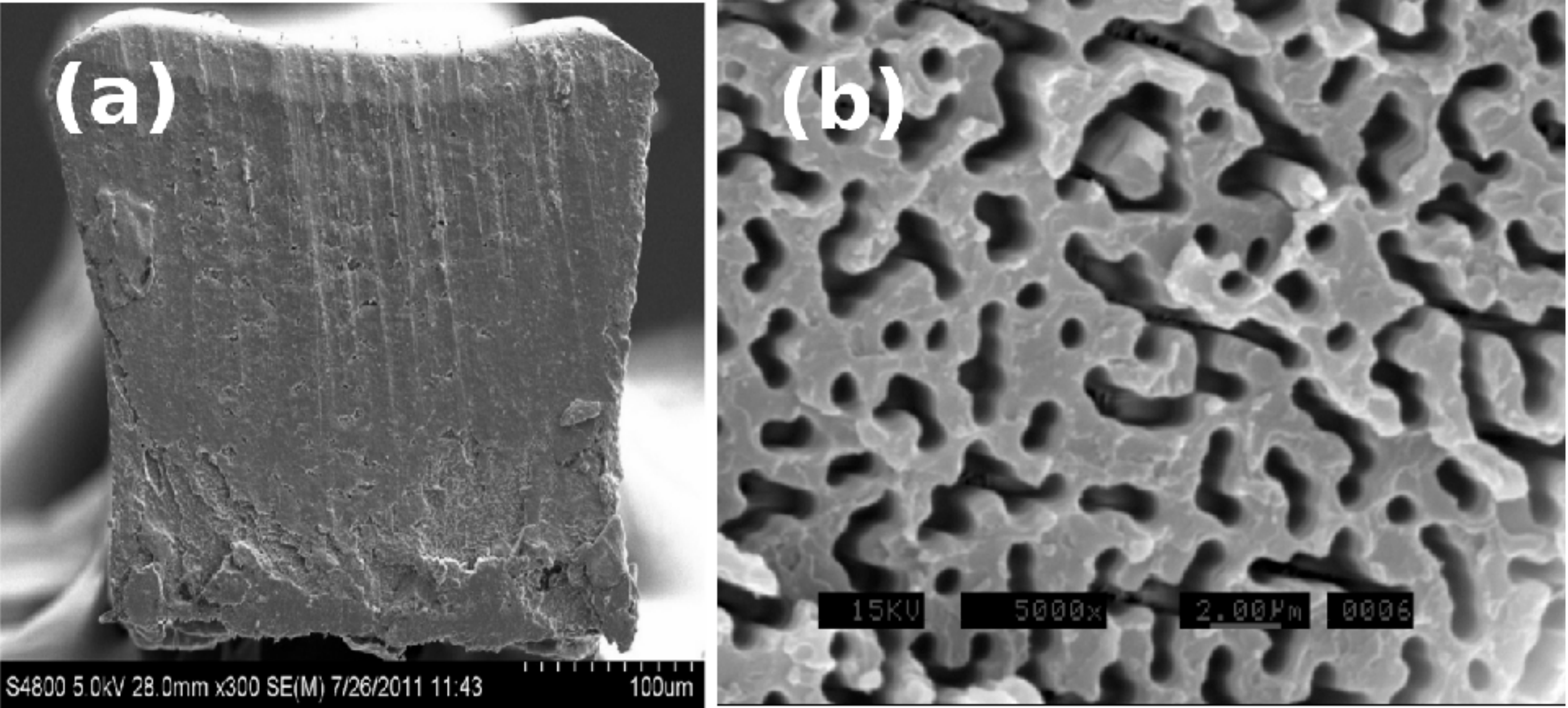}
\caption
{
(a) Cross section view of the polymer Anderson localization optical fiber developed by
Karbasi \textit {et al}.~\cite{Mafi-AL-OL-2012} with a nearly square profile and an approximate 
side width of 250~$\mu$m; (b) zoomed-in scanning electron microscope image of a 24~$\mu$m wide region on the tip
of the fiber exposed to a solvent to differentiate between PMMA and PS
polymer components, where feature sizes are around 0.9~$\mu$m and darker regions are PMMA. 
{\em Adapted with permission, copyright 2012, Optical Society of America~\cite{Mafi-AL-OL-2012}}.
}
\label{fig:SEM-AL}
\end{figure}
In 2012, Karbasi \textit {et al}.~\cite{Mafi-AL-OL-2012} reported the first observation of transverse Anderson localization
in an optical fiber. In order to obtain large refractive index fluctuations required for a strong transverse localization, 
they randomly mixed 40,000~pieces of polymethyl methacrylate (PMMA) fiber with refractive index of 1.49,
and 40,000 pieces of polystyrene (PS) fiber with refractive index of 1.59, and redrew the random stack to a 250~$\mu$m-wide optical 
fiber with random index fluctuations equal to 0.1 and an approximate transverse random feature size equal to 0.9~$\mu$m (see Fig.~\ref{fig:SEM-AL}). 
The large index contrast of 
0.1 ensured that both the beam localization radii and the sample-to-sample fluctuations over the ensemble of localized beam radii
are sufficiently small to ensure that disordered fiber operates as a genuine optical fiber--they observed that the localized beam radii launched at different transverse
positions over the facet of Anderson localizing fiber were all small and were nearly identical~\cite{Mafi-AL-OPEX-2012}. Later, 
Karbasi \textit {et al}. reported the first observation of Anderson localization in a silica optical fiber as well~\cite{Mafi-AL-OMEX-2012}. 
The reported glass-air disordered fiber was made from ``satin quartz'', which is a porous artisan glass. After drawing the preform, 
the airholes (bubbles) in the glass were stretched to form the hollow air-rods required for transverse Anderson localization. 
The large draw ratio sufficiently preserved the longitudinal invariance, without significant disturbance over typical lengths used in the experiments.
Following the observation of transverse Anderson localization in polymer~\cite{Mafi-AL-OL-2012,Mafi-AL-OPEX-2012,Mafi-AL-JOVE-2013}
and glass fibers~\cite{Mafi-AL-OMEX-2012,Mafi-AL-MDPI-2014}, 
the research team also reported the observation of beam multiplexing~\cite{Mafi-AL-OPEX-2013}, 
image transport~\cite{Mafi-AL-OC-2013,Mafi-AL-NatureComm-2014}, wave-front shaping and sharp 
focusing~\cite{Marco-AL-NatureComm-2014,Marco-AL-NatureComm-2017}, nonlocal nonlinear behavior~\cite{Marco-AL-PRL-2014,Marco-AL-APL-2014}, 
single-photon data packing~\cite{Marco-AL-ScientificRep-2016}, random lasing~\cite{Mafi-AL-LightSciApp-2017},
in addition to theoretical work on the design and optimization of Anderson localization fibers for 
image transport applications~\cite{Mafi-AL-JOSAB-2013,Mafi-AL-OL-2015,Mafi-AL-OC-2016,Mafi-AL-PRB-2016}.

\section{Transverse Anderson localization and Kerr nonlinearity in theory and experiment}

Concrete experimental results on the interplay between disorder and nonlinearity are rather sparse.
Even in the very few available published work in the literature, one does not get a clear sense on the 
extent to which Anderson localization is affected by nonlinearity. This is most likely due to the 
large parameter space and the possibility of many configurations under which the studies can be performed.
In this article, we review the results of two pioneering works by Schwartz \textit {et al}.~\cite{Schwartz2007}
and Lahini \textit {et al}.~\cite{Lahini2008}, involving both experiment and theory, and then examine 
further theoretical considerations on this issue. The interaction between disorder and other forms of 
nonlinearity (non-Kerr) will be discussed in the next section.

Earlier, we mentioned the numerical and experimental work of Schwartz \textit {et al}.~\cite{Schwartz2007} that 
resulted in the observation of transverse Anderson localization of light. The authors also investigated the
transverse Anderson localization of light in the presence of Kerr nonlinearity, both numerically and experimentally. The defining 
equation for the nonlinear propagation of light is the \NLSE~(NLSE) which includes a Kerr nonlinearity term~\cite{AgrawalBook,Mafi-NL-JLT-2012}: 
\begin{equation}
\label{eq:bpmNL}
i\dfrac{\partial A}{\partial z}+
\dfrac{1}{2n_0k_0}\left[\nabla^2_T A+k_0^2\left(n^2-n^2_0\right)A\right]+k_0 n_2 |A|^2 A=0,
\end{equation}
where $A$ is slow-varying amplitude of the optical field, $k_0$ is the vacuum wavevector, $z$ is the propagation
distance, $T$ represents the transverse coordinates ($x$ and $y$ here), $n(x,y)$ is the transversely disordered
refractive index profile, $n_0$ is the mean value of the refractive index around which index fluctuation occur, 
and $n_2$ is the nonlinear index, which is positive for self-focusing and negative for self-defocusing
nonlinearity.

As a case study, the authors considered a disordered lattice where the maximum contribution of the nonlinear 
term to the index change ``${\rm max}(|n_2|\times|A|^2)$'' was assumed to be a maximum of 15\% of the index contrast of the underlying 
periodic waveguide. They also varied the disorder level from 0\% to 30\%, where the disorder level was
defined as the magnitude of random index fluctuations relative to the index contrast of the underlying 
periodic waveguide. Using numerical simulations, they observed that over this range, the self-defocusing nonlinearity ($n_2$) results in a 
moderate (nearly negligible) widening of the average beam profile. They reported that after a short
propagation distance, the beam broadens and the remaining propagation is essentially 
as if the medium were linear, primarily because the intensity is much lower and the nonlinear effects do not play a role anymore. 
In contrast, the self-focusing nonlinearity ($n_2>0$)
resulted in a substantial reduction of the average localized beam diameter. The enhancement of localization due
to the self-focusing nonlinearity was particularly noticeable when the disorder level was less than 15\%.
They observed that at high disorder levels, where localization takes place, the difference between linear and nonlinear 
propagation is reduced, and the behavior is dominated primarily by disorder.

The experiments were performed at different nonlinear strengths. Consider $\alpha$ as the ratio between the peak intensity of 
the probe beam and the maximum intensity of the lattice-forming beams in the photorefractive crystal. The experiments were
carried out for values of $\alpha=1$, 2, and 3. The statistical analysis of the localized beam radius
clearly confirmed the expected reduction in the average beam radius due to the self-focusing nonlinearity.
They observed that self-focusing enhances localization, altering the intensity profile from diffusive-like 
to exponentially decaying. They increased the nonlinearity (increased $\alpha$) and observed that the intensity profile
narrowed down accordingly. At $\alpha=3$, the output beam profile resembles the input profile, suggesting 
the formation of a soliton.

\begin{figure}[h]
\centering\includegraphics[width=2in]{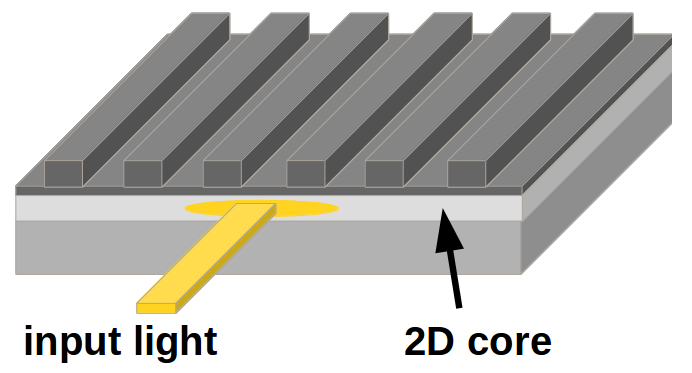}
\caption{
Schematic view of the disordered lattice waveguide used in the experiment by Lahini \textit {et al}.~\cite{Lahini2008}.
}
\label{fig:Lahini}
\end{figure}

Similar results were reported by Lahini \textit {et al}.~\cite{Lahini2008} using disordered 1D 
waveguide lattices, as shown in Fig.~\ref{fig:Lahini}. Their experiment consisted
of a 1D lattice of $N=99$ coupled optical waveguides patterned 
on an AlGaAs substrate. The \NLSE~governing the propagation of light in this coupled lattice system can be expressed as
\begin{align}
i\dfrac{\partial U_n}{\partial z} =\beta_n U_n+C(U_{n+1} + U_{n-1})+\gamma |U_n|^2 U_n,
\label{eq:coupledmodeNL0}
\end{align} 
where $n$ is the index labeling lattice sites (waveguides), $U_n$ is the optical field amplitude at site $n$, 
$\beta_n$ is the propagation constant associated with the nth waveguide, $C$ is the tunneling 
rate between adjacent sites, and $z$ is the longitudinal space coordinate. $\gamma$ characterizes the 
nonlinear Kerr effect. Disorder was introduced to the lattice by randomly changing the width of each waveguide,
resulting in the randomization of the propagation constants over a range of $\beta_0\pm\Delta$, where 
$\beta_0$ is the propagation constant for a mean value of the waveguide width. $\Delta/C$ is defined as the
disorder strength, with the implicit assumption that the coupling coefficient between adjacent waveguides is 
nearly identical. 

Light was injected into one or a few waveguides at the input, and light intensity distribution 
was measured at the output. In the linear regime, Lahini \textit {et al}. identified that highly localized eigenmodes 
exist near the top edge of the propagation constant band with a flat-phase profile and near the bottom edge of the
band with a highly varying-phase profile, having phase flips between adjacent sites. The modes near the middle of the 
band were not as localized and also showed some phase variations. 
In the weak nonlinear regime, Lahini \textit {et al} observed that nonlinearity enhances localization in
flat-phased modes and induces delocalization in the staggered modes. This behavior is explained as
follows: the presence of the weak nonlinearity perturbatively shifts (increases) the value of the 
propagation constant of each localized mode. For the flat-phased modes, the nonlinearity shifts the modes
outside the original linear spectrum. However, for the staggered, which belong to the bottom edge of the 
propagation constant band, a perturbative increase in the value of the propagation constant shifts 
it further inside the original linear spectrum. Therefore, the propagation constant of a staggered mode can 
cross and resonantly couple with other modes of the lattice, resulting in delocalization. 

The effect of nonlinear perturbations on localized eigenmodes was studied by Lahini \textit {et al}. experimentally
via exciting a pure localized mode and increasing the input beam power. The intensities were kept far below the
self-focusing threshold in the periodic lattice. They observed that some of the localized modes exhibited significant
response to nonlinearity. The experiments showed that weak positive nonlinearity tends to further localize flat phased 
localized modes, but tends to delocalize staggered modes, consistent with the theoretical prediction above.
The results obtained by Lahini \textit {et al}. do not contradict the general observations of
Schwartz \textit {et al}.~\cite{Schwartz2007}; rather, they argue that the enhanced localization due to nonlinearity
in Ref.~\cite{Schwartz2007} is only applicable to certain optical modes.

We would like to point out an early work by Pertsch \textit {et al}.~\cite{Pertsch2004} in 2004 who 
experimentally investigated light propagation in a disordered two-dimensional array of mutually
coupled optical fibers. In the linear case they observed that light either spreads in a diffusive manner or 
localizes at a few sites. The absence of Anderson localization was most likely due to the fact that the disorder
was not sufficiently high to localize the field whiting the transverse dimensions of the fiber. However, they
reported that for high excitation power diffusive spreading is arrested by the focusing nonlinearity and a 
discrete soliton is formed.
\subsection{Further theoretical considerations}

In 2008, Pikovsky and Shepelyansky presented a somewhat different account of the interaction between
disorder and nonlinearity~\cite{Pikovsky2008}. In their analysis, they considered a discrete 
\NLSE~with a third-order diagonal Kerr nonlinear term, isomorphic to Eq.~\ref{eq:coupledmodeNL0}, to study
the temporal evolution of quantum wavefunction on coupled lattices in the form of  
\begin{align}
i\dfrac{\partial \psi_n}{\partial t} =E_n\psi_n+\beta |\psi_n|^2 \psi_n+V(\psi_{n+1} + \psi_{n-1}),
\label{eq:coupledmodeNL1}
\end{align} 
where $\beta$ characterizes nonlinearity, $V$ is the hopping matrix element and is deterministic and identical for all terms. 
The disorder is introduced through the diagonal terms, where they are assumed to be randomly and uniformly distributed
in the range $-W/2<E_n<W/2$, and 
\begin{equation}
\sum_n|\psi_n|^2=1
\label{eq:norm}
\end{equation}
is assumed. For optical waveguides, this equation describes the propagation of
light in a coupled waveguide lattice, where the coupling strengths among the waveguides are identical but the propagation constants
of individual waveguides are randomized by manipulating the geometry or the refractive index of each individual waveguide.
Therefore, this conforms well with the problem studied by Lahini \textit {et al}.~\cite{Lahini2008} described earlier. 
The normalization condition means that the total optical power remains unchanged as the light propagates through the
coupled waveguide lattice. The authors verified that in the absence of nonlinearity ($\beta=0$) and in the presence of 
weak disorder, all eigenstates are exponentially localized as expected from Anderson localization theory with the localization length
$l\approx 96(V/W)^2$ at the center of the energy band~\cite{Kramer1993}.

Pikovsky and Shepelyansky considered the spreading of a field that is initially localized at the central lattice point with $|\psi_n(0)|^2=\delta_{n,0}$.
Of course, this problem is fully studied in the limit of no disorder with $W=0$ and a general value of nonlinearity $\beta\neq 0$, 
e.g., in the context of nonlinear light propagation in discrete lattices, where diffraction and nonlinearity can give rise to 
interesting physics including diffraction-free propagation and self-localized states or discrete solitons~\cite{Christodoulides2003}. 
The opposite limit of no nonlinearity with $\beta=0$ in the presence of disorder $W\neq 0$ is also well studies in the context of Anderson
localization~\cite{Kramer1993}. The intermediate regime is the focus of the work presented by Pikovsky and Shepelyansky. 

Intuitively speaking, one may think that in a nonlinear disordered coupled waveguide system, the dynamics of the
beam is initially influenced by nonlinearity; and as the beam spreads, the effect of nonlinearity becomes 
weaker and the disorder dynamics takes over. Therefore, one should always expect Anderson localization after 
a sufficiently long time. The analysis in Ref.~\cite{Pikovsky2008} is presented in view of earlier work by Sanchez-Palencia \textit {et al}.~\cite{SanchezPalencia2007}
who similarly argued for the localization of the field at large time scales, consistent with the intuitive view discussed above. 
However, Sanchez-Palencia \textit {et al}.~argued that the high-momentum cutoff of the Fourier transform of the correlation function for 1D speckle 
potentials can change localization from exponential to algebraic. Per argument presented in Ref.~~\cite{Pikovsky2008}, at first glance, one may think that 
Kerr nonlinear effects may be favored strongly by localized field configurations because if
the field spreads over $\tilde{N}$ lattice sites, the conservation of power implies that the Kerr 
nonlinearity should scale as $|\psi_n|^2\sim 1/\tilde{N}$. However, Pikovsky and Shepelyansky argue that 
the nonlinear frequency shift $\beta|\psi_n|^2\sim \beta/\tilde{N}$
should be compared with the characteristic distance between the frequencies of the exponentially localized modes, which scales as 
$1/\tilde{N}$; therefore, the effect of nonlinearity persists and does not quantitatively depend on the width of the field distribution 
and is omnipresent. 

Using a theoretical argument, they suggested that in the presence of non-vanishing nonlinearity, the beam spreading 
characterized by
\begin{equation}
\sigma^2=\sum_n(n-\langle n \rangle)^2|\psi_n|^2,
\label{Eq:2ndmoment}
\end{equation}
should follow a subdiffusive behavior where $\sigma^2\propto t^{\alpha}$
with $\alpha=2/5$. They verified their theoretical argument via numerical integration
of Eq.~\ref{eq:coupledmodeNL1} and monitoring the results, up to $t=10^8$.
For the numerical simulation they used the boundary condition $\psi_n(t=0)=\delta_{n,0}$,
where $n=0$ represents the middle waveguide, and the integration is performed by the
operator splitting method. In a sample set of simulations they chose the nonlinearity strength to be 
$\beta=V$ for two cases: case 1 with $W=2$; and case 2 with $W=4$. As expected, the initial expansion was ballistic for either case, 
but after some time $t_0$, the expansion became subdiffusive in either case. They fit the subdiffusive expansion to 
$\sigma\propto t^\alpha$ over the range $t_0<t<10^8$ to obtain an estimate for the subdiffusive exponent. 
In case 1, for different instances of randomness, they obtained
$0.32\le \alpha \le 0.39$; and for case 2 they reported $0.28\le \alpha \le 0.41$. Upon averaging 
over 8 independent realizations, they reported a fit of the form $57.5\times t^{0.344}$ for case 1 and 
$8.7\times t^{0.306}$ for case 2 over the subdiffusive ranges. 

Pikovsky and Shepelyansky analyzed the probability distribution $w_n=|\psi_n|^2$ over lattice sites $n$; in the absence of nonlinearity
they observed an exponential decay with an exponent consistent with the Anderson localization. In the presence 
of nonlinearity, however, they observed a rather flat distribution in the vicinity of $n=0$ where the width of the 
flat distribution depended on the value of $\beta$. As expected, the (low-intensity ) tails of the distribution always 
followed a decay exponent consistent with the linear theory of Anderson localization. 
Another important observation reported in Ref.~\cite{Pikovsky2008} was the presence of a certain critical strength $\beta_c$ for nonlinearity,
beyond which the reported  delocalization behavior happens. The numerical solutions suggested a value of $\beta_c\approx 0.1$
for this threshold value. However, Pikovsky and Shepelyansky argued that the threshold-like behavior might not be perfect and a slow spreading may 
persist all the way down to very small values of nonlinearity albeit at an extremely slow rate that could not be detected in 
the numerical integration scheme. 

In another study published by Kopidakis \textit {et al}.~\cite{Kopidakis2008} on the spreading of an initially localized wave packet,
the authors confirmed that while there are many initial conditions such that the second moment of the norm in Eq.~\ref{Eq:2ndmoment} 
and energy density distributions diverge with time in agreement with Ref.~\cite{Pikovsky2008}, the participation number of a wave 
packet does not diverge simultaneously. The participation number is defined as
\begin{equation}
P=\dfrac{(\sum_n|\psi_n|^2)^2}{\sum_n|\psi_n|^4},
\end{equation}
which is an alternative measure of the wave spreading to Eq.~\ref{Eq:2ndmoment}. 
Kopidakis \textit {et al}. prove this result analytically for norm-conserving models 
that satisfy Eq.~\ref{eq:norm} with strong enough nonlinearity. They showed that initially localized wave packets with a large
enough amplitude cannot spread to arbitrarily small amplitudes. The consequence is that a part of the initial energy must remain well focused at all times.

A later report by Fishman \textit {et al}.~\cite{Fishman2012} provide a somewhat different account of the impact of nonlinearity
on Anderson localization in disordered lattices. Fishman \textit {et al}.~\cite{Fishman2012} present a thorough survey of the many 
subtleties involved regarding the interaction of nonlinearity and disorder. The conclusion is that the situation can best be 
described as inconclusive at this point. For example, when using the numerical simulations, they caution that Eq.~\ref{eq:coupledmodeNL1}
is chaotic with an exponential sensitivity to numerical errors. Because of the possibly chaotic motion due to the presence of nonlinearity,
 Fishman \textit {et al}. argue that the numerical solutions of Eq.~\ref{eq:coupledmodeNL1} are not actual solutions. 
In order to draw conclusions similar to those by Pikovsky and Shepelyansky~\cite{Pikovsky2008}, one must assume that the numerical
solutions are statistically similar to the correct solutions with no real theoretical support for this assumption.
For long-time (or long-distance propagation in waveguides), it is not clear that 
reducing the $t$ step size can control the cumulative numerical error, given that the limit of zero $t$ step
may be singular. Details of these arguments can be found in Ref.~\cite{Fishman2012}.

\section{Other forms of nonlinearity}

Other forms of nonlinearity besides Kerr can also interact with the disorder-induced localization. Recently
Leonetti \textit {et al}.~\cite{Marco-AL-PRL-2014} explored the impact of thermally induced nonlinearity on
Anderson localization for a beam of light propagating in the polymer Anderson localizing optical fiber of
Ref.~\cite{Mafi-AL-OL-2012} and Fig.~\ref{fig:SEM-AL}. They reported the observation of a self-focusing action 
occurring in the disordered fiber with the binary index distribution which was triggered by a defocusing
thermal nonlinearity. The larger light absorption strength in PMMA than PS results in an inhomogeneous 
temperature distribution. The higher temperature in PMMA translates into a decrease of its refractive index. The result 
is an increased refractive index mismatch and stronger localization. In effect, they demonstrated that transversely 
localized modes shrink when the pump intensity is increased despite the fact that $n_2<0$ for the polymers,
which seems quite counter-intuitive in the first glance.

In a subsequent publication~\cite{Marco-AL-APL-2014}, 
the authors provided further evidence of this behavior by analyzing the direct relation between the optical 
intensity and the localization length. In their experiments, they probed the behavior of light by using both 
a broadband (a femtosecond Ti:Sapphire laser at 800~nm wavelength with 80~nm bandwidth) and a
monochromatic laser (solid-state continuous Nd:YAG laser at 1064 nm). They observed that the broadband light
beam injected in the fiber is dispersed at the output into a series of peaks corresponding
to localized modes. This output spectrum is modified when the input intensity in the system is increased,
demonstrating that the nonlinearity also affects the modes structure. Importantly, they found that the spatial 
extension of the individual modes decreases with fluence due to the thermal nonlinearity. They repeated
the experiment using a monochromatic continuous wave (CW) laser, activating only a single spatial mode, and
observed similar self-focusing behavior with the optical power. By comparing their observations with standard 
homogeneous polymer fibers, they proved that the disorder has turned the defocusing medium into a focusing one.

Last, we would also like to highlight another work by Karbasi \textit {et al}.~\cite{Mafi-AL-FIO-2013}, where 
they showed that the presence of disorder in transverse Anderson-localized optical waveguides enhances the nonlinear 
coefficient of the modes and creates an abundance of multimodal phase-matching opportunities, resulting in rich 
nonlinear behavior. They argued that unlike conventional multimode waveguides, in which the effective mode areas are 
typically large, resulting in reduced effective nonlinear coefficients, the localized nature of transverse Anderson-localized 
modes results in large effective nonlinear coefficients; and an abundance of phase-matching opportunities are provided due to 
the highly multimode and random nature of these waveguide.

\end{document}